\documentclass[conference]{IEEEtran}

\usepackage{float}
\usepackage{cite}
\usepackage{algorithmic}

\usepackage{algorithm}

\usepackage{amsmath}
\usepackage{amssymb}

\usepackage{listings}

\usepackage{pseudocode}

\usepackage{fancyheadings}

\usepackage{epsfig}

\usepackage{color}
\usepackage{bm}

\usepackage{amssymb, amsmath}
\DeclareMathOperator{\snr}{\mathsf{SNR}}
\DeclareMathOperator{\sinr}{\mathsf{SINR}}

\DeclareMathOperator{\expect}{\mathbb{E}}

\begin{document}
\title{End-to-End Energy Efficiency Evaluation for B5G Ultra Dense Networks\vspace{-1cm}}
\author{Yu Fu$^1$, Mohammad Dehghani Soltani$^2$, Hamada Alshaer$^2$, \\Cheng-Xiang Wang$^{1, 3, 4,*}$, Majid Safari$^2$, Stephen McLaughlin$^1$, and Harald Haas$^2$ \\%$^\dagger$
%$^\dagger$
$^1$\small{Institute of Sensors, Signals and Systems, School of Engineering and Physical Science, Heriot-Watt University, UK.}\\
$^2$\small{LiFi R$\&$D Centre, Institute for Digital Communications, School of Engineering, University of Edinburgh, UK.} \\
$^3$\small{National Mobile Communications Research Laboratory, School of Information Science and Engineering, Southeast University, Nanjing, China.}\\
$^4$\small{Purple Mountain Laboratories, Nanjing, 211111, China.}\\
$^*$\small{Corresponding author}\\
Emails:\{y.fu, cheng-xiang.wang, s.mcLaughlin\}$@$hw.ac.uk,
\{m.dehghani,h.alshaer, majid.safari, h.haas\}$@$ed.ac.uk,
chxwang@seu.edu.cn
}
\maketitle

\vspace{-0.1cm}

\begin{abstract}
Energy efficiency (EE) is a major performance metric for fifth generation (5G) and beyond 5G (B5G) wireless communication systems, especially for ultra dense networks. This paper proposes an end-to-end (e2e) power consumption model and studies the energy efficiency for a heterogeneous B5G cellular architecture that separates the indoor and outdoor communication scenarios in ultra dense networks. In this work, massive multiple-input-multiple-output (MIMO) technologies at conventional sub-6 GHz frequencies are used for long-distance outdoor communications. Light-Fidelity (LiFi) and millimeter wave (mmWave) technologies are deployed to provide a high data rate service to indoor users. Whereas, in the referenced non-separated system, the indoor users communicate with the outdoor massive MIMO macro base station directly. The performance of these two systems are evaluated and compared in terms of the total power consumption and energy efficiency. The results show that the network architecture which separates indoor and outdoor communication can support a higher data rate transmission for less energy consumption, compared to non-separate communication scenario. In addition, the results show that deploying LiFi and mmWave IAPs can enable users to transmit at a higher data rate and further improve the EE. 
\end{abstract}
\IEEEpeerreviewmaketitle
\vspace{-0.5cm}
\section{Introduction}

\begin{figure}[t!]
	\centering
	\includegraphics[width=85 mm,height=65mm]{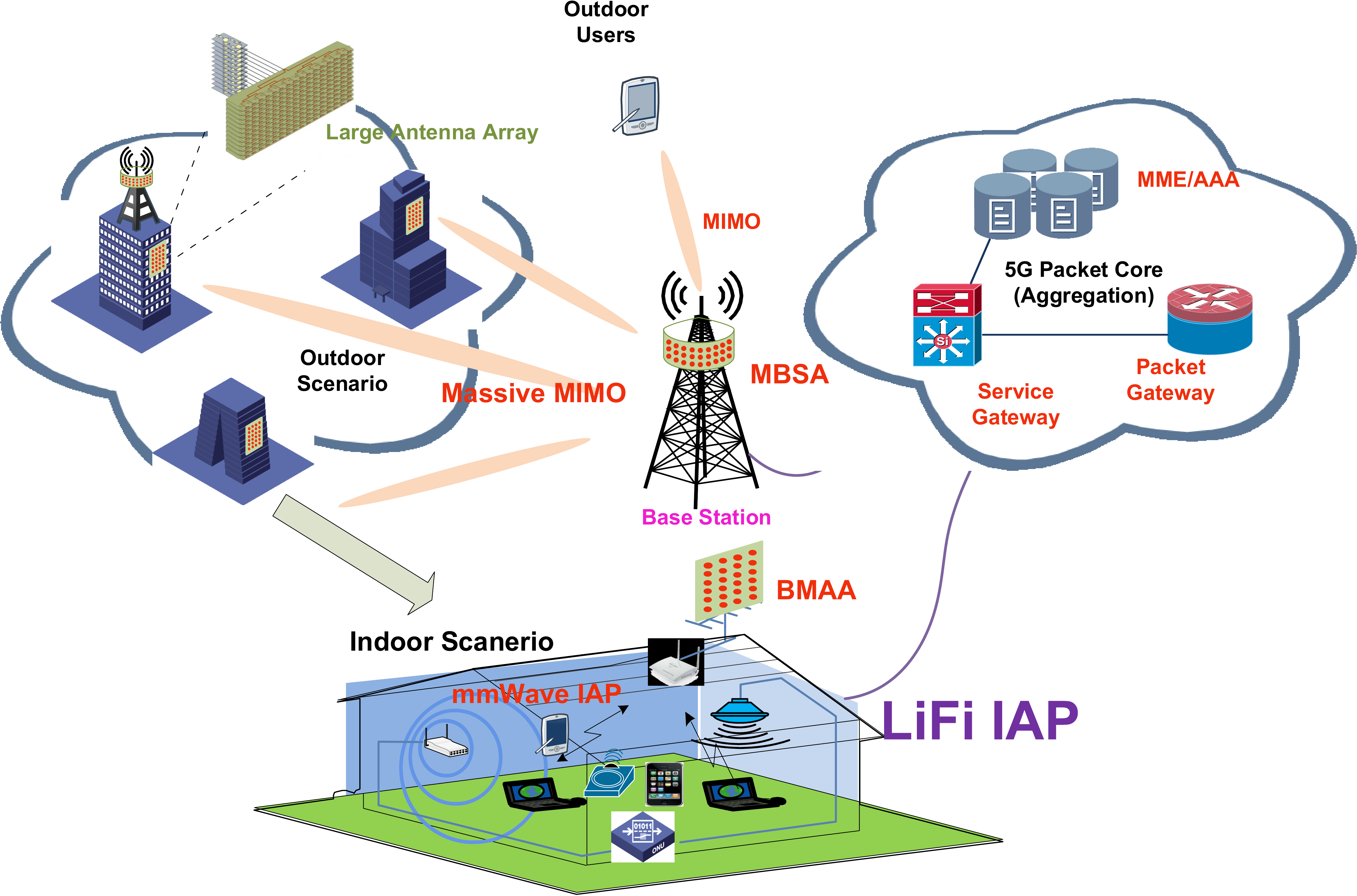}
	\vspace{-0.2cm}
	\caption{A potentional B5G network architecture.}
	\label{fig_system}
	\vspace{-0.3cm}
\end{figure}

$5$G wireless communication systems are designed to offer a significant improvement in system capacity, spectral efficiency, average cell throughput, and EE when compared with the fourth generation (4G) wireless systems. It is widely accepted that 5G and B5G network architecture will combine macrocells, picocells and small cells to support reliable, resilient and efficient wireless services for ultra dense networks~\cite{Ref:2503E}. In~\cite{Ref:2489E}, a B$5$G heterogeneous cellular architecture that can separate the outdoor and indoor communication scenarios is proposed. In this architecture, a macro base station (MBS) is assisted by the massive MIMO technology and antenna arrays (MBSALA) geographically distributed in the cell. Each antenna array serves a certain area and can be installed on an exterior wall or on the top of buildings which is referred to as the building mounted antenna array (BMAA). The BMAA is connected with indoor access points (IAPs) via fibres, as shown in Fig.~\ref{fig_system}. In this work, we considered two short-range IAP technologies, namely the beamforming based mmWave technology~\cite{MM-WveRef} and LiFi wireless technology~\cite{Ref:E2027}, for high dense indoor connections. Since the outdoor and indoor communications operate in different frequency bands, the interference between the indoor and outdoor user equipments (UEs) is avoided. Besides, the high penetration loss of mmWave signals and the small coverage for the visible light signal of LiFi also reduce the interference among the IAPs deployed in the neighboring rooms and buildings, which is helpful for building ultra dense networks. In this work, we assumed a BMAA communicates with a MBSALA using conventional Sub-6 GHz frequency band. A line of sight (LoS) path is ensured between the MBSALA and its serving BMAA. By using beamforming technologies, the LoS path can be further exploited to multiple virtual sub-beams between multiple BMAAs and a single MBSALA.
As UEs can support multiple rate access technologies (RATs) in current communication systems. In this work, we assume indoor users to communicate via their small cell IAPs in the separated condition, and directly with the MBSALA in the non-separated condition. In these 5G communication scenarios, the indoor UEs can transmit at a very high data rate, of the order Gbps, with a minimum power consumption. \\
\indent
LiFi attocell technology modulates the data through existing illumination light emitting diode (LEDs) and reuses the illumination power to support high speed optical wireless communications~\cite{Ref:2498E,MDS2019Thesis,MDS2018Orientation}. The support of multiple Gbps rates with low power consumption puts LiFi-enabled wireless communication systems among the best candidate technologies for B5G mobile wireless systems and beyond~\cite{Ref:2489E}. Some existing researches have focused on developing modulation techniques to increase the transmission data-rates~\cite{Ref:2490E} of the order Gbps at typical illumination levels in LiFi~\cite{Ref:2491E}. \\
\indent
Most existing research studies to date, on EE, have only focused on some specific technologies rather than a complete network, such as \cite{HaiderEEConnitive} for cognitive radio networks, \cite{PiyaSEMIMO} for massive MIMO systems, \cite{HaiderSEFemotocell} for mobile femtocell systems, and \cite{KuSE} for relay-sided cellular networks. References these works, this paper develops an e2e energy consumption model for evaluating the EE of a future B5G network architecture that can support ultra dense connections. In this flexible network architecture, B5G networks are established with massive MIMO considering sub-6 GHz RF frequencies, small cell mmWave, and LiFi optical wireless technologies. The developed energy consumption model covers the individual aspects of a MBS, wireless communication links and small cell APs. These are relevant for the power consumption analysis, particularly the transmission bandwidth and the number of RF chains. This study can be used as a reference for the selection of communication  technologies for different UEs, as well as the number of antennas in massive MIMO/LiFi communication systems. Compared to other existing works, this new model allows the detailed quantification of energy used by specific components, which enable a more accurate EE study at the network level. Offloading UEs from a small cell's IAP to the MBS and vice-versa is an important key in the B5G strategy vision to boost the capacity and EE of B5G wireless networks. Thus, whenever alternative connection technologies are available (as often happens in indoor scenarios), cellular traffic can be offloaded. The developed energy consumption model can be used to evaluate the total energy consumption of UEs in terms of their data transmission, which makes it suitable for investigating offloading cases.\\
\indent
The rest of this paper is organized as follows. In Sections~\ref{Sec:Outdoor}, channel models and signal propagation models of the proposed system are introduced. In Section~\ref{Massive:MIMOModel}, the power consumption model of the B5G network architecture with mmWave IAP and LiFi attocell are explained.
In Section~\ref{Simulation:Results}, the results of performance evaluation for both scenarios are discussed.
Finally, Section~\ref{Sec:Conclusion} draws conclusions of this paper.
\vspace{-0.2cm}
\section{Channel model and signal detection}
\label{Sec:Outdoor}
An outdoor massive MIMO channel which is represented by a  beamforming channel $\ell$ at MBSALA $i$ and BMAA $j$,
can be mathematically described by
\begin{equation}
\bm{G}_{\ell ji}= \sqrt{\beta_{\ell ji} M_T M_R}\bm{a}^*_{M_R}(\phi_{\ell ji})\bm{a}_{M_T}^T(\theta_{\ell ji})
\end{equation}
where $\jmath = \sqrt{-1}$, $\theta_{\ell ji}$ is the angle of arrival (AoA) at MBSALA $i$, ${M_T}$ and ${M_R}$ are the number of transmission and receiver antenna elements, respectively, $\phi_{\ell ji}$ is the
angle of arrival (AoA) at BMAA $j$, $\beta_{\ell ji}$ is the path loss, and
\begin{equation}
\bm{a}_{M_R/T}(\theta) \hspace{-0.1cm}=\hspace{-0.1cm} \frac{1}{\sqrt{M_R/T}}[ 1,\hspace{-0.1cm} ..., e^{-\jmath 2\pi \Delta (m-1)\theta}\hspace{-0.1cm},e^{-\jmath 2\pi \Delta (M_R/T-1)\theta}]^T
\end{equation}
is the antenna response of receiver or transmitter side.
$\Delta$ is the normalized antenna separation. Without loss of generality, we assume $\Delta=1/2$.
The transmission beam vector at BMAA $j$ is $\bm{a}^*_{M_T}(\theta_{\ell ji})$, and the received beam vector at BMAA $j$ is
$\bm{a}_{M_R}^T(\phi_{\ell ji})$.
$\theta_{\ell ji}$ and $\phi_{\ell ji}$ can be adjusted and optimized when deploying MBSALA and BMAA,
and thus the beamforming channel between MBSALA and BMAA can be considered static and known to both sides.\\
\indent
The downlink signal vector of MBSALA $i$, $\bm{x}_{i} $, is a linear combination of beamformed signals destined to the $N_{b}$ BMAAs,
which can be expressed as
\begin{equation}\label{out_sig}
\bm{x}_{i} = \sum_{j=1}^{N_b}\sum_{\ell=1}^{L} \bm{b}_{\ell ji} s_{\ell ji}
\end{equation}
where $s_{\ell ji}$ is an i.i.d. complex Gaussian random variable with zero mean and variance $P_{\ell ji}$,
$\bm{b}_{\ell ji} = \bm{a}^*(\theta_{\ell ji})$ denotes the beamforming vector $\ell$ to BMAA $j$. In this work, the maximum-ratio combining (MRC) precoding is applied. A perfect channel state information (CSI) is known at the MBSALA.\\
\indent
The received signal at BMAA $j$ over the beamforming channels $\ell$ is given by
\begin{align}
\label{rec_bmaa}
r_{\ell ji} =
\bm{a}^T(\phi_{\ell ji})\left(\bm{G}_{\ell ji}\bm{x}_{i} 
+\bm{n}\right)=\hspace{-0.1cm} \sqrt{\beta_{\ell ji}M_T M_R}s_{\ell ji} \hspace{-0.1cm} +{n'}
\end{align}
where $\bm{a}^T(\phi_{\ell ji})$ is the received beam vector $\ell$ at BMAA $j$,
$\bm{n}$ is the received Gaussian noise vector with $\bm{n}\sim \mathcal{CN}(\bm{0}, \sigma^2_n \bm{I}_{M_R})$.
As the link between the BMAA and MBSALA is relateive stable, the orthogonality of transmitted and received beamforming vectors enables the receivers to avoid any interference from signals sent to other buildings. \\
\indent
For the indoor mmWave link, the expression of channel model is very similar, but the interference from other indoor users should be considered. The received signal at the $k$-th UE can be given by
\begin{align}\label{rec_iue}
&r_k = \bm{g}_k^{T} \bm{b}_{m_{k}}s_k +  \sum_{j=1 j\neq k}^{N_{\mathrm{iue}}}\bm{g}_{k}^{T} \bm{b}_{m_{j}} s_{j}+  n\\\nonumber
&=\hspace{-0.1cm}\sqrt{\beta_k}F_{M'_T}(\theta_{m_k}\hspace{-0.1cm}\hspace{-0.1cm}-\hspace{-0.1cm}\theta_{k})
\hspace{-0.1cm}+\hspace{-0.1cm} \sqrt{\beta_k}\hspace{-0.1cm}\sum_{j=1 j\neq k}^{N_{\mathrm{iue}}}
F_{M'_T}(\theta_{m_{k}}\hspace{-0.1cm}-\hspace{-0.1cm}\theta_{j})s_j + n
\end{align}
where $F_{M}(x)=\frac{\sin{\pi Mx/2}}{M\sin(\pi x/2)}$ is the Sinc function. $\theta_{m_k}-\theta_k$ is the difference between the beam vector and the angle of departure(AoD). \\
\indent
For the indoor LiFi link, a desirable LoS channel gain of LiFi attocell IAP is given by \cite{Soltani2018Bi}
\begin{equation}
\label{equation1}
	H_{{\rm{LOS}}} \!=\! \begin{cases} \! \dfrac{(m+1)A}{2\pi d^{2}}\cos ^{m}\!\phi g_{\rm{f}}g(\psi)\cos\psi, & 0\le \psi\le \Psi_{c} \\
	\ 0 , & \psi _{i j}>\Psi_{\rm{c}} \end{cases}
\end{equation}
where $A$ denotes the physical area of detector, $d$ denotes the distance between the LiFi attocell IAP and  receivers (UEs), $\phi$ and $\psi$ denote the angle of radiance with respect to the axis normal to the transmitter surface, and the angle of incidence with respect to the axis normal to the receiver surface, respectively. $g_{\rm{f}}$ denotes the gain of optical filter, and $\Psi_{\rm{c}}$ denotes the field-of-view (FoV) of receiver.
Note that $g(\psi) =\varsigma^2 /\sin^2\Psi_{\rm{c}}$, for $0\le\psi_i\le\Psi_{\rm{c}}$, and $0$ for $\psi_{i}>\Psi_{\rm{c}}$, represents the
optical concentrator gain, where $\varsigma$ denotes the refractive index. The Lambertian order is obtained
from $m = -1/\log_2(\cos\Phi_{1/2})$, where $\Phi_{1/2}$ is the half-intensity angle~\cite{ChenCheng}.\\
\indent
The radiance angle, $\phi$, and the incidence angle, $\psi$, of the LiFi IAP and the receiver are given based on the analytical geometry rules, such
as $\cos\phi={\bf{d}}\cdot{\bf{n}}_{\rm{tx}}/{\Vert {\bf{d}}\Vert }$ , and $\cos\psi=-{\bf{d}}\cdot{\bf{n}}_{\rm{rx}}/{\Vert {\bf{d}}\Vert }$, where ${\bf{n}}_{\rm{tx}}=[0, 0, -1]$ and ${\bf{n}}_{\rm{rx}}=[0, 0, 1]$ are the normal vectors at the LiFi IAP and the receiver planes, respectively.
And ${\bf{d}}$ denotes the distance vector between the LiFi AP and the receiver, and $\Vert \cdot\Vert$ and $``\cdot"$ stand for the Euclidean norm operators
and the inner product, respectively~\cite{ChenCheng}. For more information about 5G channel models, please refer to \cite{5Gchannel,5GchannelmmWave,5GchannelmmWave2,WuTCOM18}.
\vspace{-0.2cm}
\section{Energy efficiency}
\label{Ener:Model}
Energy efficiency, $\eta$, measures the effectiveness of converting power into data
traffic transmission. It is defined as the spectral efficiency divided by the total power consumption, given by
\vspace{-0.2cm}
\begin{equation}
\eta = \frac{\mathbb{C}}{P}.
\end{equation}
The spectral efficiency is defined by the Shannon equation as
\begin{equation}
\mathbb{C}= \gamma\expect\left[\log_2(1+\sinr)\right]
\end{equation}
where $\sinr$ is the signal-to-interference-noise-ratio under the given channel, and $\gamma$ is the channel usage efficiency factor.
Based on the approximation derived in \cite{mMIMOScaling}, for a massive MIMO system,
we assume $\expect[\log_2(1+\sinr)i]\approx \log_2(1+\expect[\sinr])$,
the spectral efficiency analysis can be reduced to the analysis
of the expectation of SNR, $\expect[\snr]$.\\
\indent
Based on Eq.(\ref{rec_bmaa}), the expectation of received SNR over a link $\ell$ between
the MBSALA $i$ and BMAA $j$ can be expressed as
\vspace{-0.2cm}
\begin{align}\label{SINROUT}
\expect\left[\snr_{\ell ji}^{\mathrm{M} \rightarrow \mathrm{B}}\right] =\frac{{\beta_{\ell ji}}M_T P_{\ell ji}}
{\sigma^2_n/M_R}
\end{align}
where $P_{\ell ji}$ denotes the signal variance.
Samilar to \ref{SINROUT} For the indoor mmWave link, its SINR can be expressed as
\begin{equation}\label{sinr_iue}
\expect\left[\sinr_k^{\mathrm{I}\rightarrow \mathrm{U}} \right]\vspace{-0.1cm}=\vspace{-0.1cm}
\frac{\beta_k \expect\left[F_M^2(\theta_k-\theta_{m_k})\right]P_k} 
{ \beta_k \expect\left[\sum_{j=1, j\neq k}^{N_{\mathrm{iue}}}F_M^2(\theta_k\vspace{-0.1cm}-\vspace{-0.1cm}\theta_{m_j})\right]  P_j \vspace{-0.1cm}+ \vspace{-0.1cm}\sigma_n^2}.
\end{equation}
For the LiFi link, the SINR is given by:
\begin{equation}
\sinr=\frac{c_{f}^{2} P_{\rm{t}}^2H_{\rm{LOS}}^2}{  N_0B  +  \sum\limits_{j=1}^{N_{F}}   c_{Ijf}^{2}  P_{I\rm{j}}^2H_{Ij\rm{LOS}}^2}
\end{equation}
where $N_{F}$ denotes the number of interfering LiFi attocell IAPs, $I$ denotes the symbol for interfering LiFi attocell IAPs,
$c_{f}$ denotes the LED coefficient~\cite{Ref:2703E}, $P_{t}$ denotes the optical transmitted power,
$N_{0}$ denotes the noise spectral density.
\subsection{System power consumption model}
\label{Massive:MIMOModel}
The total power consumption of the system is the sum of the power consumed by all the power devices
deployed in a wireless communication cell.
In the separate outdoor and indoor scenario, the total power consumed by MBS, MBSALA, BMAA
and IAP can be expressed as
\begin{equation}
P_\mathrm{cell} = P^{\mathrm{MBS}}_{\mathrm{tot}} + \sum_{i=1}^{N_a}(P^{\mathrm{BMAA}}_{\mathrm{tot}}+P^{\mathrm{IAP}}_{\mathrm{tot}}).
\end{equation}
where $N_{a}$ denotes the number of MBSALA.
A high level energy efficiency evaluation framework (E\textsuperscript{3}F) for mobile communication systems was investigated and developed by the Energy Aware Radio and network technologies (EARTH) project \cite{Ref:2488E}. Based on the framework, the general energy consumption model of a mobile communication system is given by
\vspace{-0.2cm}
\begin{equation}
P_{\mathrm{sys}} = P_{\mathrm{BB}} + P_{\mathrm{RF}} + P_{\mathrm{PA}} + P_{\mathrm{OH}}
\end{equation}
where $P_{\mathrm{BB}}$ represents the power consumed by the digital baseband processing unit, $P_{\mathrm{RF}}$ and $P_{\mathrm{RF}}$
represent the power consumption of RF front-end (FE) and power amplifier (PA), and $P_{\mathrm{OH}}$ represents the power overhead.
This is mainly consumed by the system cooling unit and AC-DC converters.\\
\indent
The digital baseband processing includes digital signal processing, and system control and network processing.
For digital signal processing, the operations include digital filtering, up/down sampling, (I)FFT, MIMO channel training
and estimation, OFDM modulation/demodulation, symbol mapping and channel encoding/decoding.
The operation complexity denoted by $\mathcal{O}$, is measured by Giga floating-point operations per second (GFO/S), depending
on the type of the operation and number of UEs and data streams. The power consumption per Giga floating-point operation is further
scaled by a technology-dependent factor $\rho$. Thus, $P_{\mathrm{BB}}$ can be obtained by dividing GFO/S by $\rho$. For systems nowadays,
the factor is $\rho = 160$ GOP/W.\\
\indent
The key constituent components of RF FE include carrier modulators, frequency synthesis, clock generators,
digital to analogue /analogue to digital converters, mixers and so on. The power consumption of these components
scale with parameters, namely system bandwidth, number of antennas and traffic load.
The power consumption model of PA depends on the type and the maximum output power of the amplifier. It
is also related to the actual output power that assures the desired spectral efficiency.
This paper considers two types of power amplifiers: Class-B PA and Doherty PA.
The class-B PA is equipped in the MBSALA and BMAA, which enables relatively high output power transmission.
The Doherty PA is developed for high-frequency band communication systems
with high power efficiency. It is deployed in the IAPs. The power models for MBS, MBSALA, BMAA and IAP are developed in the
following subsections.
% Explicit power models for MBS, BMAA and IAP are introduced in the following subsections.
% The power models for MBS, MBSALA, BMAA and IAP are developed, respectively.
% In the following sections,
% the expected SINR of three types of constituent links in separate outdoor and indoor scenario are analyzed
\subsubsection{Power consumption model of MBS}
The total power of MBS is given by
\begin{equation}
P^{\mathrm{MBS}}_{\mathrm{tot}} = \frac{
P^{\mathrm{MBS}}_{\mathrm{BB}} +  {N_a}P^{\mathrm{MBSALA}}_{\mathrm{tot}} }
{(1-\eta_{\mathrm{c}})(1 - \eta_{\mathrm{ac-dc}})(1 - \eta_{\mathrm{dc-dc}})}
\end{equation}
where $P^{\mathrm{MBS}}_{\mathrm{BB}}$ is the power of  digital base band processing at MBS, $P^{\mathrm{MBSALA}}_{\mathrm{tot}}$ is the total power consumed by a MBSALA, and $\eta_{\mathrm{c}}$, $\eta_{\mathrm{ac-dc}}$, $\eta_{\mathrm{dc-dc}}$ are the power efficiency of the cooling system, AC-DC and DC-DC conversion, respectively.\\
\indent
Based on the outdoor distributed antenna architecture, the digital baseband processing at MBS includes mapping/de-mapping of symbols, channel encoding ($\mathcal{O}_{\mathrm{enc}}$), upper layer network ($\mathcal{O}_{\mathrm{nw}}$) and control operations ($\mathcal{O}_\mathrm{ctrl} $) for the $N_a$ MBSALA,
respectively. Thus, $P^{\mathrm{MBS}}_{\mathrm{BB}}$ can be further written by
\vspace{-0.5cm}
\begin{equation}
P^{\mathrm{MBS}}_{\mathrm{BB}} = \sum_{i = 1}^{N_a}(
\mathcal{O}_{\mathrm{ctrl},i} +\mathcal{O}_{\mathrm{network},i}+ \mathcal{O}_{\mathrm{enc},i})/\rho.
\end{equation}
\subsubsection{Power consumption model of MBSALA}
The power consumption of MBSALA can be decomposed as
\begin{equation}
P^{\mathrm{MBSALA}}_{\mathrm{tot}} =
P^{\mathrm{MBSALA}}_{\mathrm{BB}} + P^{\mathrm{MBSALA}}_{\mathrm{RF}}+
P^{\mathrm{MBSALA}}_{\mathrm{PA}}.
\end{equation}
For MBSALA, the downlink baseband processes include filtering, up sampling, (I)FFT of OFDM symbols, massive
MIMO channel estimation, precoding/beamforming, symbol mapping, and control and network relation operations,
which can be expressed as
\begin{align}
\label{MBSALA_bb}
&P^{\mathrm{MBSALA}}_{\mathrm{BB}} = \\ \nonumber
&(\mathcal{O}_{\mathrm{fltr}} \hspace{-0.1cm}+ \hspace{-0.1cm}\mathcal{O}_{\mathrm{fft}} \hspace{-0.1cm} + \hspace{-0.1cm}\mathcal{O}_{\mathrm{est}} \hspace{-0.1cm}+ \hspace{-0.1cm} \mathcal{O}_{\mathrm{bf}} \hspace{-0.1cm}+ \hspace{-0.1cm}\mathcal{O}_{\mathrm{pre}} \hspace{-0.1cm} + \hspace{-0.1cm} \mathcal{O}_{\mathrm{map}}  \hspace{-0.1cm}+ \hspace{-0.1cm} \mathcal{O}_{\mathrm{ctrl}} \hspace{-0.1cm}+  \hspace{-0.1cm}\mathcal{O}_{\mathrm{nw}})/\rho.
\end{align}
Here the complexity of (I)FFT can be scaled by $\mathcal{O}_{\mathrm{(i)fft},i} = N_{\mathrm{s}}N_{\mathrm{(i)fft}}\log_2(N_{\mathrm{(i)fft}})$,
where $N_{\mathrm{(i)fft}}$ is the number of sub-carriers of an OFDM symbol and $N_{\mathrm{s}}$ is the total number of OFDM symbols.
The complexity of channel estimation by correlation of orthogonal pilot sequences can be scaled by $\mathcal{O}_{\mathrm{est}}=\tau M_T N_{\mathrm{ue}}$,
where $N_{\mathrm{ue}}$ denotes the number of UEs, and  we let $\tau = N_{\mathrm{ue}}$.
The complexity of channel precoding and beamforming operations can be scaled by $\mathcal{O}_{\mathrm{bf/pre}} = (N_{\mathrm{ue}} + N_bL)(1 - \tau/N_{\mathrm{c}})$ for uplink channel estimation.
%For the operation complexity of the remaining processes, we will refer to the typical values provided in~\cite{modellte}.
The RF FE (front-end) of MBSALA has  modulator,  mixer, clock generation and D-A converter. The total power consumption
of RF FE (front-end) is given by
\begin{eqnarray}
P^{\mathrm{MBSALA}}_{\mathrm{RF}}= M_T(P_{\mathrm{mod}}^{\mathrm{MBSALA}} +  P_{\mathrm{mix}}^{\mathrm{MBSALA}} + \\ \nonumber P_{\mathrm{dac}}^{\mathrm{MBSALA}})
+    \sqrt{M_T}P_{\mathrm{clk}}^{\mathrm{MBSALA}}.
\end{eqnarray}
The power consumption of  the modulator, mixer and DAC scales linearly with the number of antennas. The power consumption of clock generator scales by the
square root of the number of antennas.
\subsubsection{Power consumption model of BMAA}
The power consumption model of BMAA can be expressed as follows:
\begin{equation}
P^{\mathrm{BMAA}} = \frac{
P^{\mathrm{BMAA}}_{\mathrm{BB},\ell} +  M_RP^{\mathrm{BMAA}}_{\mathrm{RF}} }
{(1-\eta_{\mathrm{c}})(1 - \eta_{\mathrm{ac-dc}})(1 - \eta_{\mathrm{dc-dc}})}.
\end{equation}
In this work, we assume a BMAA receives the downlink traffic from its serving MBSALA over $L$ beamforming data links and forwards it to an IAP.
The baseband data processes consist of filtering, beamforming process, sampling, IFFT process, symbol de-mapping, channel
decoding, control and network processes. The GOP/S can be estimated similarly to the case of MBSALA. The power consumption of baseband processing in
BMAA can be given as follows:
\begin{align}
\label{bmaa_bb}
&P^{\mathrm{BMAA}}_{\mathrm{BB}} =  \\ \nonumber
&L \cdot(\mathcal{O}_{\mathrm{fltr}}  \hspace{-0.1cm} + \hspace{-0.1cm} \mathcal{O}_{\mathrm{bf}}  \hspace{-0.1cm}+ \hspace{-0.1cm}\mathcal{O}_{\mathrm{smpl}} \hspace{-0.1cm}+ \hspace{-0.1cm}
\mathcal{O}_{\mathrm{ifft}}  \hspace{-0.1cm}+ \hspace{-0.1cm} \mathcal{O}_{\mathrm{demap}}  \hspace{-0.1cm}+ \hspace{-0.1cm} \mathcal{O}_{\mathrm{dec}} \hspace{-0.1cm}+ \hspace{-0.1cm}\mathcal{O}_{\mathrm{ctrl}}
 \hspace{-0.1cm}+ \hspace{-0.1cm} \mathcal{O}_{\mathrm{nw}})/\rho.
\end{align}
The analogue components of the downlink RF FE include mixer, clock, variable gain
amplifiers (VGA), ADC and low-noise amplifier (LNA). The power consumption of RF in BMAA can be given as follows:
\begin{eqnarray}
\label{bmaa_rf}
P^{\mathrm{BMAA}}_{\mathrm{RF}} = M_R\cdot\hspace{-0.1cm}(
\mathcal{O}_{\mathrm{mix}}\hspace{-0.1cm}+\hspace{-0.1cm}\mathcal{O}_{\mathrm{vga}} \hspace{-0.1cm}+\hspace{-0.1cm}\mathcal{O}_{\mathrm{adc}}+
\mathcal{O}_{\mathrm{lna}})\hspace{-0.1cm}+\hspace{-0.1cm}\sqrt{M_R}\mathcal{O}_{\mathrm{clc}}.
\end{eqnarray}
Note that the power consumption of LNA and VGA over the receiver RF FE are considered
constant, unlike the PA used for signal transmission.
% Energy model consumption model  (Power model distribution)
\subsubsection{Power consumption Model of mmWave IAP}
For the mmWave IAP, its power consumption model can be expressed as
\vspace{-0.2cm}
\begin{equation}
P^{\mathrm{IAP}}_{\mathrm{tot}} = \frac{
P^{\mathrm{IAP}}_{\mathrm{BB}} +  {M'_T}(P^{\mathrm{IAP}}_{\mathrm{RF}} + 
P^{\mathrm{IAP}}_{\mathrm{PA}})}
{(1 - \eta_{\mathrm{c}})(1 - \eta_{\mathrm{ac-dc}})(1 - \eta_{\mathrm{dc-dc}})}
\end{equation} 
where
\vspace{-0.2cm}
\begin{equation}
P^{\mathrm{IAP}}_{\mathrm{RF}}= M'_T(P_{\mathrm{mix}}^{\mathrm{IAP}}+P_{\mathrm{dac}}^{\mathrm{IAP}}+P_{\mathrm{bft}}^{\mathrm{IAP}}
+P_{\mathrm{fs}}^{\mathrm{IAP}})+ \sqrt{M'_T}P_{\mathrm{clc}}^{\mathrm{IAP}}
\end{equation}
and
\vspace{-0.2cm}
\begin{equation}
P_{\mathrm{PA}}^{\mathrm{IAP}} = 
\begin{cases}
\frac{2}{\pi}\sqrt{P_{\mathrm{o}} P_{\mathrm{max}}^{\mathrm{IAP}}}, \quad 0< P_{o}<0.25P_{\mathrm{max}}^{\mathrm{IAP}}\\
\frac{6}{\pi}\sqrt{P_{\mathrm{o}} P_{\mathrm{max}}^{\mathrm{IAP}}}, \quad 0.25P_{\mathrm{max}}^{\mathrm{IAP}}\leq P_{o}\leq P_{\mathrm{max}}^{\mathrm{IAP}} 
\end{cases}.
\end{equation} 
\vspace{-0.5cm}
\subsubsection{Power consumption Model of LiFi IAP}
\label{LiFi:Models}
A LiFi attocellular network consists of several small attocells. Each IAP
covers an area with radius $3-10$ m. The power consumed for lighting the
off-the-shelf LEDs in the IAP is used to support visible light communication~\cite{Ref:E2027}.
The light photons are modulated at a very high speed to support
$3.5$Gbps $@$2m distance as well as $1.11$Gbps $@$10m, with a total optical output
power of $5$ $mW$~\cite{Ref:2485E,Ref:2486E}. This provides a significant spectrum, which can support hungry bandwidth applications and emerging services with low power consumption.

\begin{figure}
	\centering
	\includegraphics[width=80mm,height=50mm]{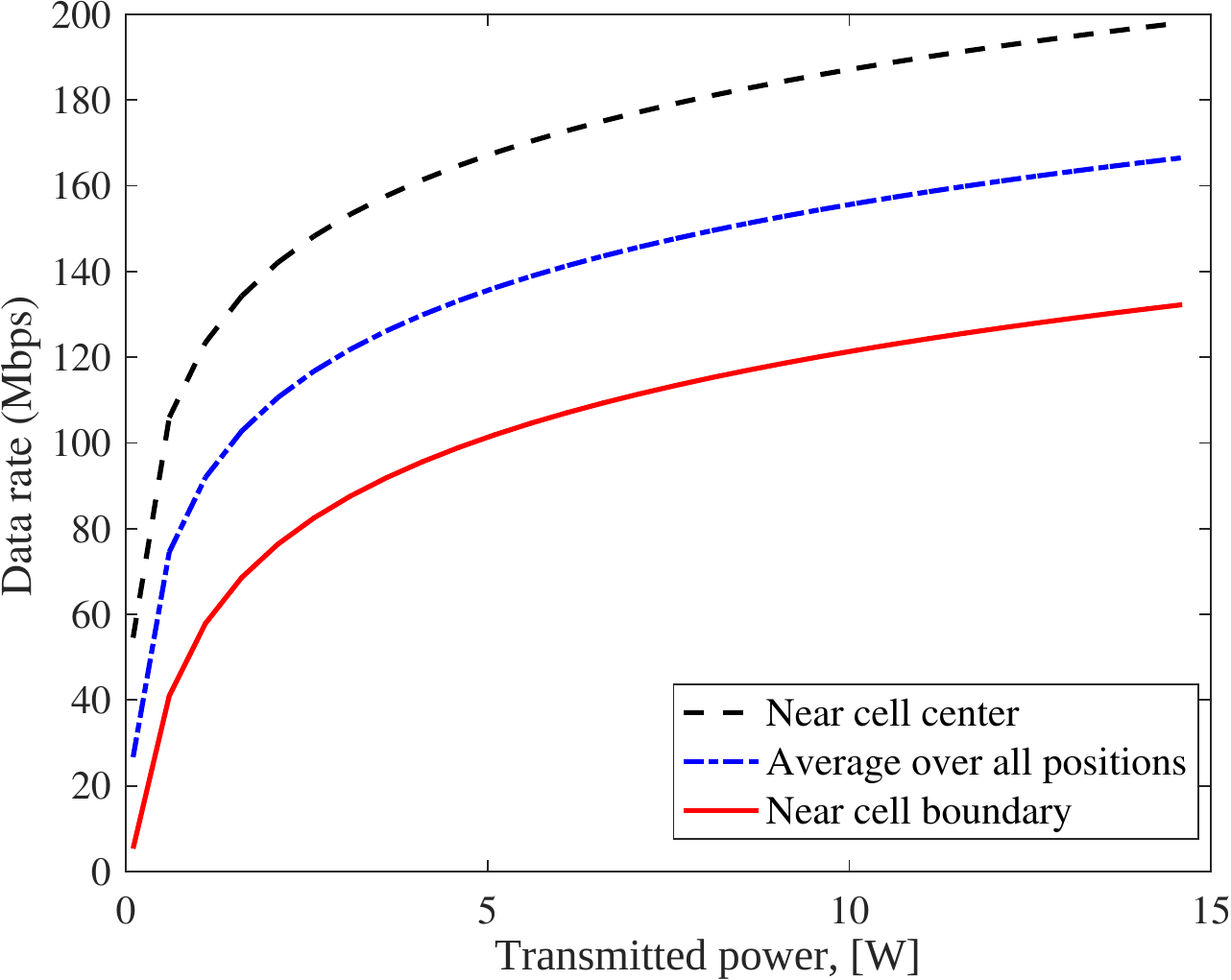}
	\vspace{-0.2cm}
	\caption{LiFi Downlink throughput distribution Vs. LiFi IAP transmitted power.}
	\label{pow:LiFi}
	\vspace{-0.3cm}
\end{figure}

The maximum number of transmitted bits per joule of input energy in a LiFi communication system is known as the
energy consumption factor (CF)~\cite{Ref:2494E}. The total power consumption in the LiFi attocellular system comprises
two main parts: the circuit (illumination) power consumption and the power consumed for transmitting the wireless data
at high data rates~\cite{Ref:2703E}. The illumination power can be expressed as follows~\cite{Ref:2703E}:
\vspace{-0.2cm}
\begin{equation}
P_{Light} =  \frac{nq V_{T} \Phi}{p_{f} \epsilon}  \ln (\frac{q\Phi}{p_{f} \epsilon I_{s}}+1);
\label{Eq:Energy}
\end{equation}
And the extra power consumed for transmission high data rates can be expressed as follows~\cite{Ref:2703E}:
\vspace{-0.2cm}
\begin{equation}
P_{comm} =  \frac{nq V_{T} H_{LOS}^{2}}{2 p_{f} \epsilon  \mu_{\Phi}};
\label{Eq:Energy1}
\end{equation}
where the variables in Eq.~(\ref{Eq:Energy}) and Eq.~(\ref{Eq:Energy1}) are defined in Eqs.(3,5) in \cite{Ref:2703E}.
Fig.~\ref{pow:LiFi} shows the throughput distribution vs. the transmitted power across a LiFi attocell.

\vspace{-0.2cm}
\section{Performance Results}
\label{Simulation:Results}
\begin{figure}
	\centering
	\resizebox{0.87\linewidth}{!}{
	\includegraphics{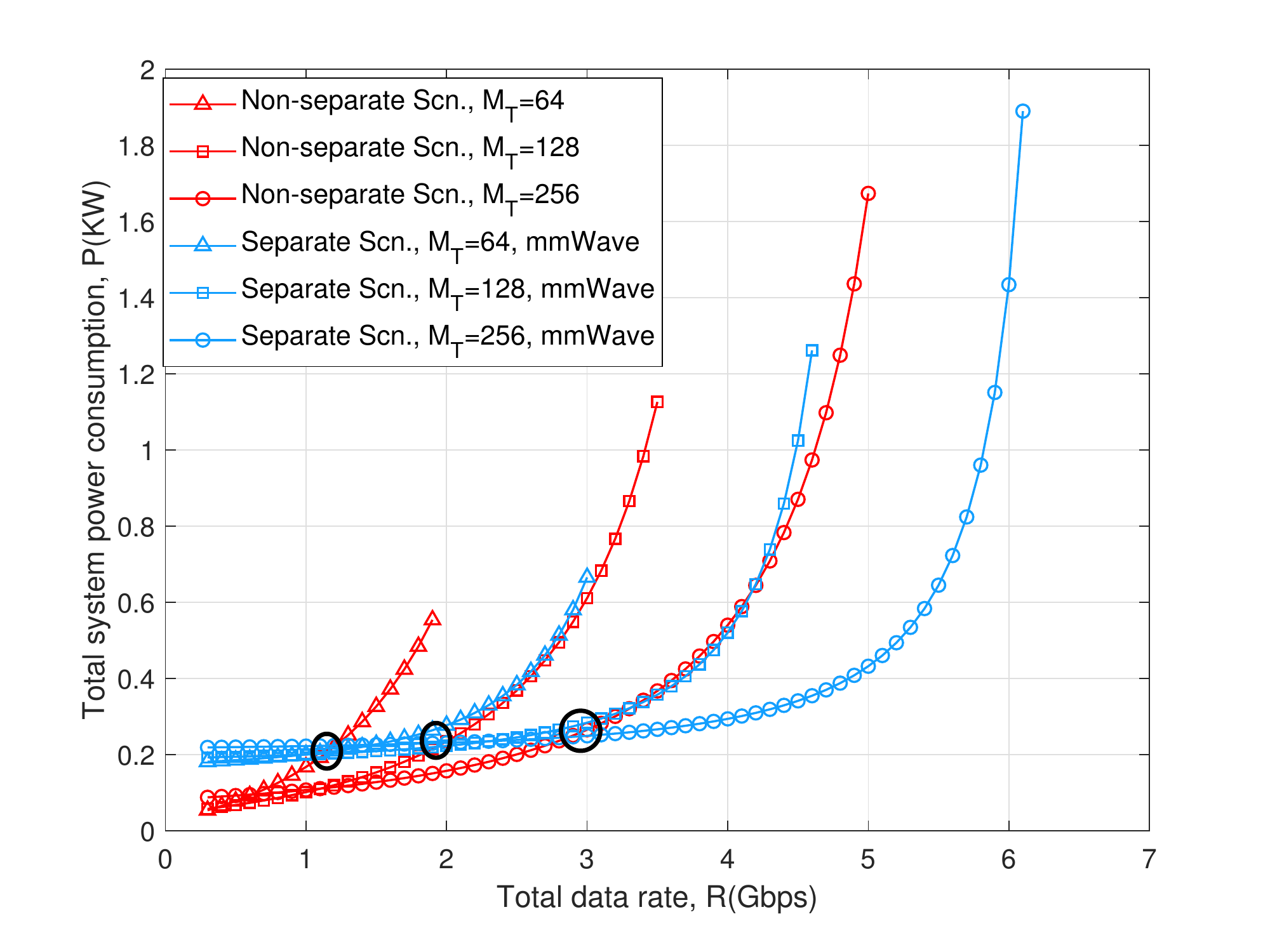}}
	\vspace{-0.3cm}
	\caption{Comparison of system power consumption for non-separate and separate scenarios with mmWave IAPs.}
	\label{pow_consumption}
	\vspace{-0.2cm}
\end{figure}

\begin{figure}
	\centering
	\resizebox{0.87\linewidth}{!}{
	\includegraphics{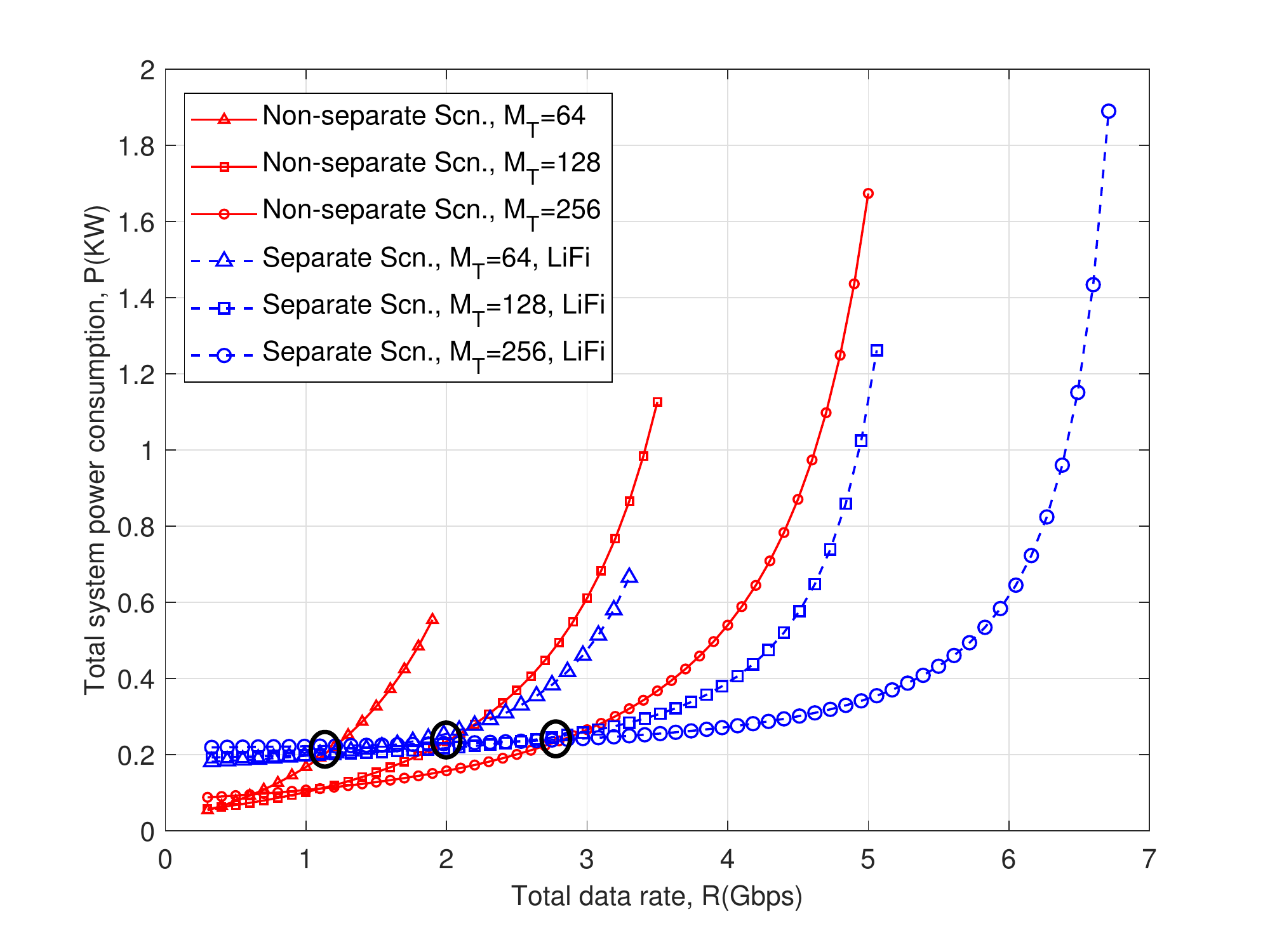}}
	\vspace{-0.3cm}
	\caption{Comparison of system power consumption for non-separate and separate scenarios with LiFi IAPs.}
	\label{EELiFi}
	\vspace{-0.3cm}
\end{figure}

\begin{figure}
	\centering
	\resizebox{0.87\linewidth}{!}{
	\includegraphics{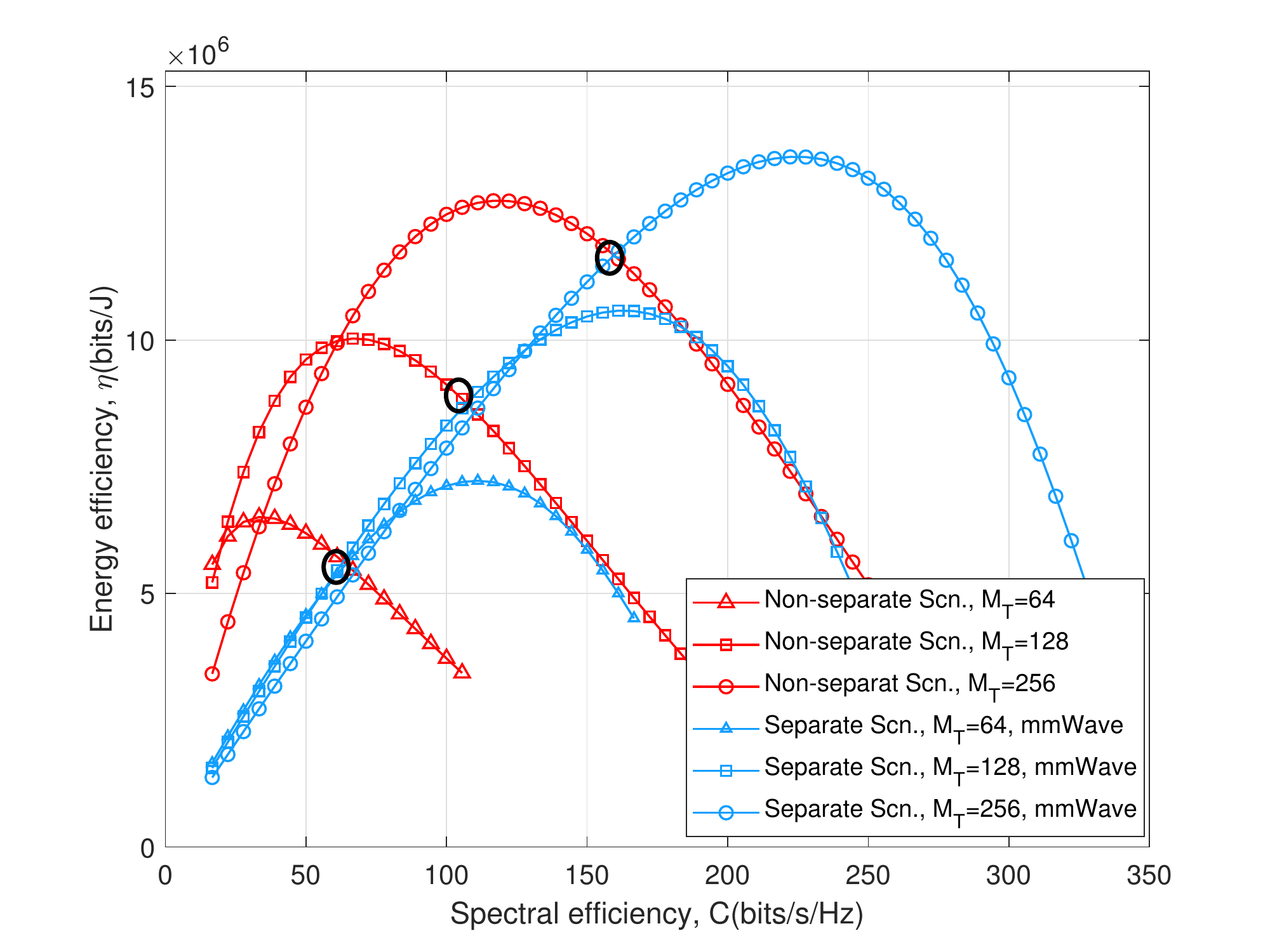}}
	\vspace{-0.3cm}
	\caption{EE and SE trade off curves for separate and non-separate scenarios with mmWave IAPs.}
	\label{ee_se}
	\vspace{-0.3cm}
\end{figure}

\begin{figure}
	\centering
	\resizebox{0.87\linewidth}{!}{
	\includegraphics{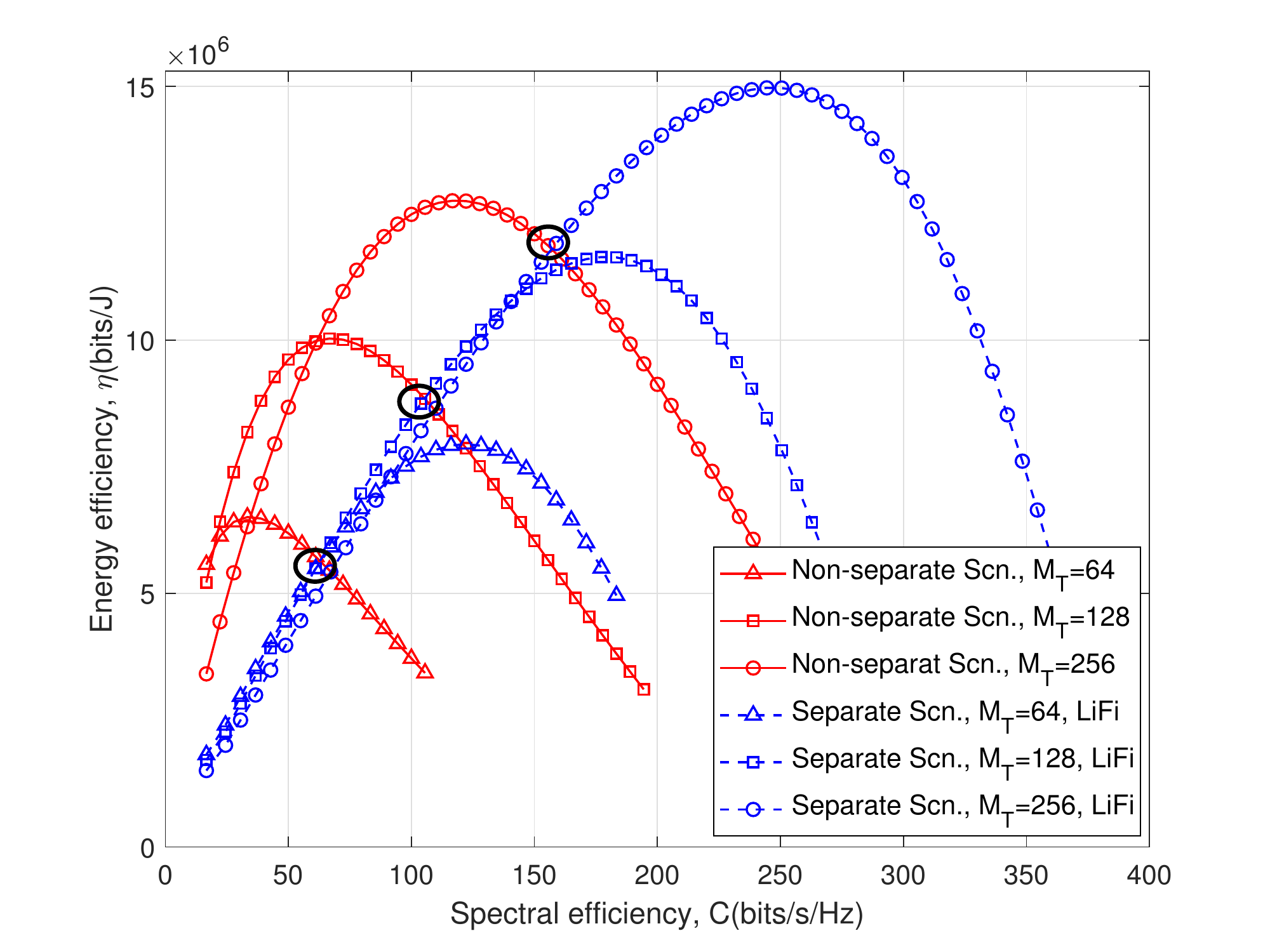}}
	\vspace{-0.3cm}
	\caption{EE and SE trade off curves for separate and non-separate scenarios with LiFi IAPs.}
	\label{pt_LiFi}
	\vspace{-0.3cm}
\end{figure}

% A simulation environment has been developed in MATLAB to compare the performance of
%The performance of the proposed separate and non-separate 5G communication scenarios are evaluated using
%MATLAB in terms of total power consumption and energy efficiency.
The performance of the proposed B5G network architecture is evaluated
and compared using MATLAB in terms of  total power consumption and energy efficiency for LiFi and  mmWave IAPs.
The performance evaluation
parameters for the mmWave AP are based on the reference~\cite{MM-WveRef}.
It is considered that each MBSALA serves $4$ buildings.
Each MBSALA is equipped with a number of antennas, which is taken to be
$(64,128, 256)$ in the separate and non-separate evaluation scenarios. Foe each BMAA, 64 antennas are equipped.
For each MBSALA and BMAA pair, $4$ beamforming links are established.
The WINNER II~\cite{Ref:E2031} B5a Rooftop (Eq.(4.23), Page 43)- Rooftop model is
adopted as the path loss model for the link between the MBSALA and BMAA, which
is given by
\vspace{-0.2cm}
\begin{equation}
PL_{\mathrm{MBSALA =}}23.5\log_{10}(d) + 42.5 + 20\log_{10}(f_c/5.0).
\end{equation}
For the non-separate coverage system, the indoor users are served directly by the MBSA and a penetration loss of 20 dB is considered.\\
\indent
Simulation results in Fig.~\ref{pow_consumption} shows the total end-to-end power consumption of the total data
rate transmission in separate and non-separate scenarios. A mmWave IAP is considered with the configuration parameters used in~\cite{MM-WveRef}.
From this figure, it can be observed that the increase in the number of antennas, $M_T$, provides an opportunity
to support more users and hence data rate transmission. The total
power consumption increases in terms of data rate transmission in both separate
and non-separate scenarios. However, at relatively low data rate transmission, the total power
consumption in separate and non-separate scenarios remains steady, despite the
increase of $M_T$. For example, the total power consumption
in the non-separate scenario remains steady until $1$ Gbps, when $M_T$ increases
from $64$ to $128$. Similarly, the total power consumption in the
separate scenario remains steady until $3$ Gbps, when $M_T$ increases
from $64$ to $128$. This can be attributed to that fact that when $M_T$
increases, the antennas and IAPs are more likely to become saturated.
For a given $M_T$, it is observed that the power consumption
curves of the two scenarios cross at some point.
On the right side of the crossing point, the total power consumption in the
separate scenario consumes less power than in the non-separate scenario to
transmit the same data rate. But on the left side of the crossing point,
this trend is in the opposite direction.
Fig.~\ref{pow_consumption} shows that more total power is consumed in the separate scenario
than in the non-separate scenario at a low data rate. But, the results show noticeable opposite
trends when the data rate is significantly increased. This is attributed to the
BMAA and IAPs deployed in the separate scenario, which consume less power to transmit
a high data rates. For example, when $M_t$=256, data rate= 5 Gbps, the total power consumption of the non-separate scenarios is 1.6 KW, while for the separated case, its consumed power is only 0.4 KW. In this case, it is recommended that the non-separate scenario is deployed. However,
an IAP LiFi attocell that serves indoor users has reduced the total energy
consumption at low and high data rates transmission by almost $10\%$ compared with a mmWave IAP,
as shown in Fig.~\ref{EELiFi}.
The separate scenario supports great data rate transmission with less power consumption, particulary in
the case of deploying  LiFi IAPs for serving UEs, as shown in Fig.~\ref{pow_consumption} and Fig.~\ref{EELiFi}.
Nevertheless, all the result trends show an increase in the power consumption when the
total transmitted data rate exceeds a specific threshold value.
This is because the power amplifiers reach their saturation point
after the total power consumption exceeds a certain level.\\
\indent
The EE versus SE trade-off curves of the two scenarios are shown in Fig.~\ref{ee_se}
and Fig.~\ref{pt_LiFi}. The trends of EE firstly ascend as SE increases
and start to descend after a certain point. For EE, it shows that the increase in the number of antennas does not change the energy
efficiency at low SE due to the extra power consumed by BMAA and IAP. However, the EE increases sharply at high SE when the number
of antennas increases. From these figures, it is clear that by separating the indoor and outdoor communication scenarios, the system can obtain a
better trade-off between SE and EE. This demonstrates, in  turn, the advantage of the proposed 5G network and communication architecture.
\vspace{-0.3cm}
\section{Conclusions}
\label{Sec:Conclusion}
This paper has introduced an e2e power consumption model for a B5G network architecture which integrated massive MIMO, indoor mmWave, and LiFi attocell technologies to fulfill the requirement of ultra dense networks. The obtained results show that by separating the indoor and outdoor communication scenarios, the B5G system can obtain a better trade-off between EE and SE. This indicates the advantage of the proposed B5G network architecture. The obtained results also indicates that the indoor UEs may be need to communicate directly to the outdoor MBS when their transmit data rate is relatively low, to achieve a better EE performance. In contrast, the UEs can communicate via the LiFi and mmWave IAPs, when their data rate is high. The general trends of performance metrics in the obtained results also confirm the performance requirement of B5G to support greater data rates, in the order of
multiple Gbps, with low power consumption and higher energy efficiency. This means that the UEs may require some metric indicators to guide them to communicate via different technologies. This can be a design issue for controlling traffic offloading from cellular networks to small cell IAPs and vice-versa.
The results also show that deploying LiFi attocell IAPs can reduce the total power
consumption by almost $10\%$ compared to the RF mmWave indoor wireless small cell technology, which shows the potential advance of LiFi. 

\section*{Acknowledgment}
\small{
This work was supported by the National Key R\&D Program of China under Grant 2018YFB1801101, the National Natural Science Foundation of China (NSFC) under Grant 61960206006, the EPSRC TOUCAN project under Grant EP/L020009/1, the High Level Innovation and Entrepreneurial Talent Introduction Program in Jiangsu, the Research Fund of National Mobile Communications Research Laboratory, Southeast University, under Grant 2020B01, the Fundamental Research Funds for the Central Universities under Grant 2242019R30001, the Huawei Cooperation Project, and the EU H2020 RISE TESTBED2 project under Grant 872172.}
\bibliographystyle{IEEEtran}
\vspace{-0.2cm}
\bibliography{Ref}

% Generated by IEEEtran.bst, version: 1.14 (2015/08/26)
\begin{thebibliography}{10}
\providecommand{\url}[1]{#1}
\csname url@samestyle\endcsname
\providecommand{\newblock}{\relax}
\providecommand{\bibinfo}[2]{#2}
\providecommand{\BIBentrySTDinterwordspacing}{\spaceskip=0pt\relax}
\providecommand{\BIBentryALTinterwordstretchfactor}{4}
\providecommand{\BIBentryALTinterwordspacing}{\spaceskip=\fontdimen2\font plus
\BIBentryALTinterwordstretchfactor\fontdimen3\font minus
  \fontdimen4\font\relax}
\providecommand{\BIBforeignlanguage}[2]{{%
\expandafter\ifx\csname l@#1\endcsname\relax
\typeout{** WARNING: IEEEtran.bst: No hyphenation pattern has been}%
\typeout{** loaded for the language `#1'. Using the pattern for}%
\typeout{** the default language instead.}%
\else
\language=\csname l@#1\endcsname
\fi
#2}}
\providecommand{\BIBdecl}{\relax}
\BIBdecl

\bibitem{Ref:2503E}
J.~G. Andrews, S.~Buzzi, W.~Choi, S.~V. Hanly, A.~Lozano, A.~C.~K. Soong, and
  J.~C. Zhang, ``{What will 5G be?}'' \emph{IEEE J. Sel. Areas Commun},
  vol.~32, no.~6, pp. 1065--1082, June 2014.

\bibitem{Ref:2489E}
C.~X. Wang, F.~Haider, X.~Gao, X.~H. You, Y.~Yang, D.~Yuan, H.~M. Aggoune,
  H.~Haas, S.~Fletcher, and E.~Hepsaydir, ``{Cellular architecture and key
  technologies for 5G wireless communication networks},'' \emph{IEEE Commun.
  Mag.}, vol.~52, no.~2, pp. 122--130, Feb. 2014.

\bibitem{MM-WveRef}
C.~Lin and G.~Y. Li, ``{Energy-efficient design of indoor mmWave and sub-THz
  systems with antenna arrays},'' \emph{IEEE Trans. Wireless Commun.}, vol.~15,
  no.~7, pp. 4660--4672, July 2016.

\bibitem{Ref:E2027}
H.~Haas, ``High-speed wireless networking using visible light,'' \emph{SPIE
  Newsroom}, Apr. 2013.

\bibitem{Ref:2498E}
S.~Rajbhandari, H.~Chun, G.~Faulkner, and K.~C. et~al., ``{High-speed
  integrated visible light communication system: Device constraints and design
  considerations},'' \emph{IEEE J. Sel. Areas Commun}, vol.~33, no.~9, pp.
  1750--1757, Nov./Dec. 2015.

\bibitem{MDS2019Thesis}
M.~D. Soltani, ``{Analysis of Random Orientation and User Mobility in LiFi
  Networks},'' \emph{The University of Edinburgh}, 2019.

\bibitem{MDS2018Orientation}
M.~D. Soltani, A.~A. Purwita, Z.~Zeng, H.~Haas, and M.~Safari, ``{Modeling the
  Random Orientation of Mobile Devices: Measurement, Analysis and {LiFi} Use
  Case},'' \emph{IEEE Trans. Commun.}, vol.~67, no.~3, pp. 2157--2172, Mar.
  2019.

\bibitem{Ref:2490E}
S.~Dimitrov and H.~Haas, ``{Information rate of OFDM-based optical wireless
  communication systems with nonlinear distortion},'' \emph{IEEE J. Lightw.
  Technol.}, vol.~31, no.~6, pp. 918--929, Mar. 2013.

\bibitem{Ref:2491E}
A.~M. Khalid, G.~Cossu, R.~Corsini, P.~Choudhury, and E.~Ciaramella, ``{1-Gb/s
  transmission over a phosphorescent white LED by using rate-adaptive discrete
  multitone modulation},'' \emph{IEEE Photon. J}, vol.~4, no.~5, pp.
  1465--1473, Oct. 2012.

\bibitem{HaiderEEConnitive}
F.~Haider, C.~X. Wang, H.~Haas, E.~Hepsaydir, X.~Ge, and D.~Yuan, ``Spectral
  and energy efficiency analysis for cognitive radio networks,'' \emph{IEEE
  Trans. Wireless Commun.}, vol.~14, no.~6, pp. 2969--2980, June 2015.

\bibitem{PiyaSEMIMO}
P.~Patcharamaneepakorn, S.~Wu, C.~X. Wang, e.~H.~M.~Aggoune, M.~M. Alwakeel,
  X.~Ge, and M.~D. Renzo, ``{Spectral, energy, and economic efficiency of 5G
  multicell massive MIMO systems with generalized spatial modulation},''
  \emph{IEEE Trans. Veh. Technol.}, vol.~65, no.~12, pp. 9715--9731, Dec. 2016.

\bibitem{HaiderSEFemotocell}
F.~Haider, C.~X. Wang, B.~Ai, H.~Haas, and E.~Hepsaydir, ``{Spectral/energy
  efficiency tradeoff of cellular systems with mobile femtocell deployment},''
  \emph{IEEE Trans. Veh. Technol.}, vol.~65, no.~5, pp. 3389--3400, May 2016.

\bibitem{KuSE}
I.~Ku, C.~X. Wang, and J.~Thompson, ``{Spectral-energy efficiency tradeoff in
  relay-sided cellular networks},'' \emph{IEEE Trans. Wireless Commun.},
  vol.~12, no.~10, pp. 4970--4982, Oct. 2013.

\bibitem{Soltani2018Bi}
M.~{D. Soltani}, X.~{Wu}, M.~{Safari}, and H.~{Haas}, ``Bidirectional user
  throughput maximization based on feedback reduction in {LiFi} networks,''
  \emph{IEEE Trans. Commun.}, vol.~66, no.~7, pp. 3172--3186, July 2018.

\bibitem{ChenCheng}
C.~Chen, D.~Basnayaka, and H.~Haas, ``{Downlink performance of optical attocell
  networks},'' \emph{J. Lightw. Technol.}, vol.~34, no.~1, pp. 137--156, Jan.
  2016.

\bibitem{5Gchannel}
C.~X. {Wang}, J.~{Bian}, J.~{Sun}, W.~{Zhang}, and M.~{Zhang}, ``A survey of
  {5G} channel measurements and models,'' \emph{IEEE Commun. Surveys \&
  Tutorials}, vol.~20, no.~4, pp. 3142--3168, Fourthquarter 2018.

\bibitem{5GchannelmmWave}
C.~X. Wang, S.~Wu, L.~Bai, X.~You, J.~Wang, and C.-L. I, ``Recent advances and
  future challenges for massive mimo channel measurements and models,''
  \emph{Sci. China Inf. Sci}, vol.~59, no.~2, pp. 1--16, Feb. 2016.

\bibitem{5GchannelmmWave2}
J.~{Huang}, C.~{Wang}, R.~{Feng}, J.~{Sun}, W.~{Zhang}, and Y.~{Yang},
  ``Multi-frequency mmwave massive mimo channel measurements and
  characterization for {5G} wireless communication systems,'' \emph{IEEE J.
  Sel. Areas Commun.}, vol.~35, no.~7, pp. 1591--1605, July 2017.

\bibitem{WuTCOM18}
S.~{Wu}, C.~{Wang}, e.~M.~{Aggoune}, M.~M. {Alwakeel}, and X.~{You}, ``A
  general {3-D} non-stationary {5G} wireless channel model,'' \emph{IEEE Trans.
  Commun.}, vol.~66, no.~7, pp. 3065--3078, July 2018.

\bibitem{mMIMOScaling}
Q.~Zhang, S.~Jin, K.~K. Wong, H.~Zhu, and M.~Matthaiou, ``{Power scaling of
  uplink massive MIMO systems with arbitrary-rank channel Means},'' \emph{IEEE
  J. Sel. Topics Signal Process.}, vol.~8, no.~5, pp. 966--981, Oct 2014.

\bibitem{Ref:2703E}
A.~Tsiatmas, F.~M.~J. Willems, and J.-P.~M. Linnartz, ``{Joint illumination and
  visible-light communication systems: data rates and extra power
  consumption},'' in \emph{{Proc. IEEE ICCW 2015}}, June 2015, pp. 1380--1386.

\bibitem{Ref:2488E}
G.~Auer, V.~Giannini, C.~Desset, and I.~Godor, ``{How much energy is needed to
  run a wireless network?}'' \emph{IEEE Commun. Mag.}, vol.~18, pp. 40--49,
  Oct. 2011.

\bibitem{Ref:2485E}
H.~Haas, ``{The future of wireless light communication},'' in \emph{{Proc. of
  2nd Int. Workshop Visible Light Commun. Syst}}, 2015, pp. 1--1.

\bibitem{Ref:2486E}
J.~Kim, C.~Lee, and J.-K.~K. Rhee, ``{Traffic off-balancing algorithm for
  energy efficient networks},'' in \emph{{Proc. of SPIE/OSA/IEEE Asia Commun.
  Photon}}, 2011, pp. 83--100.

\bibitem{Ref:2494E}
J.~N. Murdock and T.~S. Rappaport, ``{Consumption factor and power- efficiency
  factor: a theory for evaluating the energy efficiency of cascaded
  communication systems},'' \emph{IEEE J. Sel. Areas Commun}, vol.~32, no.~2,
  pp. 221--236, Feb. 2014.

\bibitem{Ref:E2031}
{IST-4-027756 WINNER II}, ``{WINNER II Channel Models},'' \emph{Deliverable WP1
  Channel Model}, pp. 1--82, Sept. 2007.

\end{thebibliography}
\end{document}